\documentclass[prl,
reprint,superscriptaddress,notitlepage,floatfix]{revtex4-2}

\usepackage{graphicx}
\graphicspath{{figs/}}
\usepackage{dcolumn}
\usepackage{bm}
\usepackage[svgnames]{xcolor}
\usepackage{soul}

\usepackage{ bbold }
\usepackage{amssymb}
\usepackage{ dsfont }
\usepackage[version=4,arrows=pgf-filled,
textfontname=sffamily,
mathfontname=mathsf]{mhchem}
\usepackage{csquotes}
\usepackage{booktabs}
\usepackage[bookmarks=false]{hyperref}
\usepackage{multirow}
\hypersetup{
    colorlinks,
    linkcolor={Crimson},
    citecolor={Green},
    urlcolor={NavyBlue}
}

\newcommand{\cO}{\mathcal{O}}
\newcommand{\cS}{\mathcal{S}}
\newcommand{\cG}{\mathcal{G}}

\newcommand{\cR}{\mathcal{R}}
\newcommand{\cL}{\mathcal{L}}

\newcommand{\bx}{\mathbf{x}}

\newcommand{\bnab}{\boldsymbol{\nabla}}

\newcommand{\C}[2]{\begin{minipage}[t]{#1}\raggedright #2\end{minipage}}

\begin{document}

\title{Perturbative criteria for the relevance of nonreciprocity on phase transitions}

    \date{\today}

\author{Giulia Garcia Lorenzana}
\affiliation{NSF-Simons National Institute for Theory and Mathematics in Biology, Chicago IL}
\affiliation{Laboratoire de Physique de l'\'Ecole normale sup\'erieure, ENS, Universit\'e PSL, CNRS, Sorbonne Universit\'e, Universit\'e de Paris F-75005 Paris, France}
\affiliation{Laboratoire Matière et Systèmes Complexes (MSC), Université Paris Cité, CNRS, 75013 Paris, France}

\author{David Martin}
\affiliation{LPTMC, Sorbonne Universit\'e \& CNRS,
Paris, 75252, France}

\author{Yael Avni}
\affiliation{James Franck Institute, University of Chicago, Chicago, IL 60637, USA}

\author{Daniel~S.~Seara}
\affiliation{Mechanical and Industrial Engineering, University of Illinois Chicago, IL 60607, USA}

\author{Michel Fruchart}
\affiliation{Gulliver, CNRS, ESPCI Paris, Université PSL, 75005 Paris, France}

\author{Giulio Biroli}
\affiliation{Laboratoire de Physique de l'\'Ecole normale sup\'erieure, ENS, Universit\'e PSL, CNRS, Sorbonne Universit\'e, Universit\'e de Paris F-75005 Paris, France}

\author{Vincenzo Vitelli}
\affiliation{James Franck Institute, University of Chicago, Chicago, IL 60637, USA}
\affiliation{Leinweber Institute for Theoretical Physics, University of Chicago, Chicago, IL 60637, USA}

    \begin{abstract}
        Nonreciprocal interactions are widely observed in nonequilibrium systems, from biological or sociological dynamics to open quantum systems.
        Despite the ubiquity of nonreciprocity, its impact on phase transitions is not fully understood.
        In this work, we derive criteria to perturbatively assess whether nonreciprocity changes the universality class {of two-species} systems undergoing a phase transition.
        These criteria, stated in terms of the unperturbed critical exponents in the spirit of the Harris criterion for disordered systems, assess whether static critical exponents change at first order under a given perturbation.
        For example, in the case of a nonreciprocal version of model A with two species, a homogeneous nonreciprocal perturbation is relevant whenever the two parts are initially identical and uncoupled, and irrelevant otherwise.
        Our results agree with existing renormalization group calculations and with numerical simulations.
    \end{abstract}
    \maketitle

In nonequilibrium systems, microscopic components can interact in a \textit{nonreciprocal} way: the effect of A on B need not be equal to the one of B on A.
Microscopic nonreciprocal interactions break time-reversal symmetry and can lead to drastic macroscopic consequences, such as the existence of nonequilibrium dynamical phases and phase transitions. However, it is also possible that they get washed out at large scales~\cite{OByrne2022,Fodor2016,Fodor2022,Granek2024,Evans1998,Sieberer2025,Daviet2024,Zelle2024,sompolinsky1986temporal}.
For instance, adding random nonreciprocal interactions in a spherical Sherrington-Kirkpatrick model destroys its spin glass phase but does not lead to time-dependent behavior, while structured nonreciprocity between two populations morphs it into an oscillating spin glass~\cite{garcialorenzana2024a,crisanti1987}.
Yet, the precise conditions under which microscopic nonreciprocity leads to observable features at macroscopic scales are still unknown.

In this Letter, we propose a procedure to perturbatively assess the effect of nonreciprocal perturbations on the universality class of systems undergoing a phase transition.
The result are a set of criteria on the critical exponents of the unperturbed system, in the spirit of the Harris criterion \cite{Harris1974,Brooks2016} for equilibrium disordered systems (row 1 in Table~\ref{figtab:schematic}).
The procedure is particularly effective in systems composed of two asymmetrically coupled fields, which have emerged as a paradigmatic way to introduce nonreciprocity across scales \cite{garcialorenzana2024a,Dinelli2023,Ivlev2015,Hong2011,Young2020,Young2024,Ott2008,Avni2025,Avni2025a,guislain2023,guislain2024c,fruchart2021,You2020,Saha2020,Duan2025, pisegna_emergent_2024}.
Examples range from predator-prey dynamics \cite{Murray1993,Tsyganov2003} and excitatory-inhibitory neuronal circuits \cite{montbrio2018kuramoto}, to open quantum systems \cite{Chiacchio2023,Nadolny2025,Jachinowski2025} and socially-driven human dynamics \cite{Seara2025,Zakine2024,garnier-brun2025}.

The scope of our Letter is not to perform detailed renormalization group (RG) calculations of critical exponents, but rather to assess whether or not a perturbation affects the static critical properties of a phase transition.
This is related, but not equivalent, to studying the relevance of the perturbation under RG flow: an RG-relevant perturbation could leave the static exponents unchanged but affect the dynamical exponents.
Our procedure, formulated within the formalism of stochastic path integrals, encompasses both equilibrium and nonequilibrium systems.

\begin{figure}
    \centering
    \includegraphics[width=0.9\linewidth]{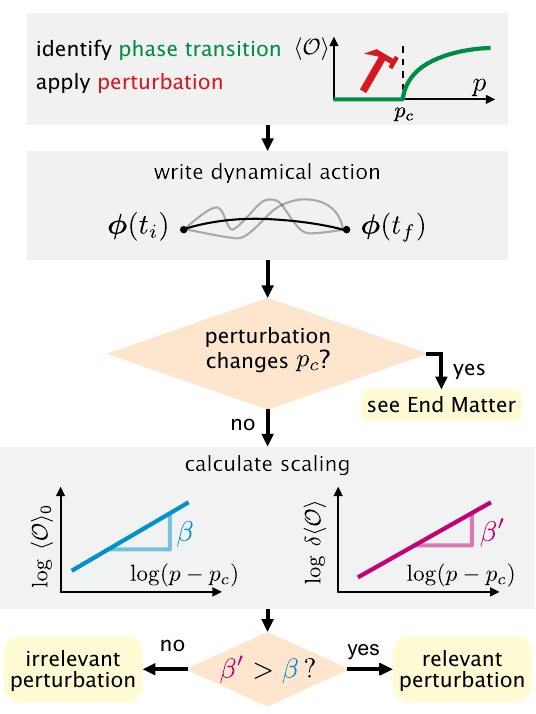}
    \caption{
    \textbf{When does a perturbation change a phase transition?}
    To answer this question, we proceed as follows.
    (i) Start with a system described by a -- potentially nonequilibrium -- action $\mathcal{S}_0$ and exhibiting a phase transition with order parameter $\mathcal{O}$ at a critical parameter $p_c$.
    (ii) Choose  a (potentially nonreciprocal) perturbation, encoded as $\delta\mathcal{S}$ in the action.
    (iii) Check if the perturbation changes the critical parameter $p_c$.
    Using extra symmetries, it can be guaranteed that this does not happen at first order (End Matter).
    When it does, a correction has to be applied so that the corrected perturbation effectively moves the system parallel to the critical line, see End Matter.
    (iv) Calculate how the perturbation to the order parameter scales near the critical point ($\delta \langle \mathcal{O} \rangle \sim (p-p_c)^{\beta'}$) and compare it to the scaling of the unperturbed observable ($\langle \mathcal{O} \rangle_0 \sim (p-p_c)^{\beta}$).
    (v) Conclude: the perturbation is relevant (i.e. changes the transition)  \st{at tree level} when $\beta' > \beta$.
    (In the cases considered in the main text, $p$ is the temperature $T$.)
    }
    \label{fig:pipeline}
\end{figure}

\textit{Nonreciprocal Model A ---}
We illustrate our approach on a nonreciprocal version of Model A, in the classification of Hohenberg and Halperin \cite{hohenberg1977}, defined by the dynamical equations
\begin{align}
\label{eq:basic_coupled_equation_1}
\begin{split}
    \partial_t\phi_1&= - V^{\prime}(\phi_1) +\nabla^2\phi_1  + [K_+ + K_-] \phi_2 + h_1 + \eta_1\\
    \partial_t\phi_2&=- V^{\prime}(\phi_2) + \nabla^2\phi_2   + [K_+ -  K_-] \phi_1 + h_2 + \eta_2
\end{split}
\end{align}
for two real-valued scalar fields $\phi_i$ ($i=1,2$), where $V(\phi) = - a \, \phi^2/2 + b  \,\phi^4/4$ is a symmetric double-well potential \footnote{Note that the $\phi_i$'s usually correspond to the coarse-grained value of a microscopic quantity: as such the coefficients of $V(\phi)$ depend on the microscopic parameters of the underlying model. Taking the example of the Ising Model, $a$ and $b$ will be functions of the coupling constant $J$.}, $h_1$ and $h_2$ are auxiliary fields (used to define response functions and otherwise set to zero),
$\eta_i(\bx,t)$ are Gaussian white noises satisfying $\langle \eta_i(\bx,t) \eta_j(\bx',t') \rangle = 2T \delta_{ij} \delta(\bx-\bx') \delta(t-t')$.
The coefficients $K_+$ and $K_-$ characterize the strength of the symmetric and antisymmetric (nonreciprocal) couplings, respectively.

When $K_+ = K_- = 0$, Eq.~\eqref{eq:basic_coupled_equation_1} describes two identical and uncoupled fields that (for spatial dimension $d \geq 2$) undergo a spontaneous $\mathbb{Z}_2$ symmetry breaking
in the Ising universality class.
We perturb this uncoupled case with a small antisymmetric interaction of strength $\delta K_-$.
When $\delta K_- \neq 0$ and $K_+ = 0$, the coupled system is still invariant under simultaneous inversion of both fields $\bm{\phi} \to - \bm{\phi}$.
To assess whether the perturbation $\delta K_-$  modifies the critical properties, we now compute the correction to the order parameters $\phi_1$ and $\phi_2$  to first order in $\delta K_-$ and compare its scaling to the unperturbed order parameter at the critical point.
If the correction can asymptotically be neglected when approaching the critical point, the perturbation is irrelevant; otherwise it can alter the critical behavior (see Fig.~\ref{fig:pipeline}).

Note that generic perturbations (such as reciprocal coupling) usually shift the critical point, even when they do not alter the universality class. This shiftleads to a trivial correction to the order parameter, which has to be subtracted before making conclusions on the relevance of the perturbation. We discuss this procedure in the EM.
Here, such shift must be at least quadratic in $\delta K_-$, since reversing $\delta K_-$ is equivalent to exchanging the identities of the two identical fields; this greatly simplifies the analysis.

\textit{Nonequilibrium dynamical action formalism ---}
The probability of observing a given configuration $\bm{\phi} = \left( \phi_1(\bx, t), \phi_2(\bx, t) \right)$ can be expressed as \cite{martin1973, dedominicis1978, janssen1976}
\begin{align}
\let\vec\bm
\mathcal{P}[\vec{\phi}]=\int D\hat{\vec{\phi}} \ e^{-\int d\bx dt \ \cS[\vec{\phi}, \hat{\vec{\phi}}]}
\end{align}
where $\hat{\phi}_i$ are auxiliary response fields. The action $\cS$ can be decomposed as $\cS=\cS_0 + \delta \cS$, where $\cS_0$ is the unperturbed action and
\begin{align}
    \label{eq:action_interaction_part}
    \delta\cS = \delta K_- (\hat{\phi}_1\phi_2-\hat{\phi}_2\phi_1).
\end{align}
is the perturbation (see SM for more details).
The average magnetization of the first field is then given by
\begin{equation}
\begin{aligned}
\!\!
    \langle \phi_1\rangle = \int \!\!D\bm{\phi} \, \phi_1 \mathcal{P}[\bm{\phi}] \simeq \textcolor[HTML]{0092C8}{ \langle \phi_1 \rangle_0} + \textcolor[HTML]{C40075}{\delta \langle \phi_1 \rangle},
\end{aligned}
\end{equation}
where $\textcolor[HTML]{0092C8}{\langle \phi_1 \rangle_0}$ represents averaging of the field with respect to $\mathcal{S}_0$, and $\textcolor[HTML]{C40075}{\delta \langle \phi_1 \rangle}$ is the first order correction due to $\delta \mathcal{S}$ (colors match Fig.~\ref{fig:pipeline}).
Expanding the exponential of the action, the correction can be written as
\begin{align}
\delta \langle\phi_1(\bx,t)\rangle = - \left\langle \phi_1(\bx,t) \int \delta \cS|_{\bx', t'} d\bx' dt'  \right\rangle_0\;.
\end{align}
Using Eq. \eqref{eq:action_interaction_part} (see SM for details), we further obtain
\begin{align}
\label{eq:nonrec_deviation}
\delta\langle\phi_1(\bx,t)\rangle =- \delta K_-\left\langle \phi_1(\bx,t) \int_{\bx', t'} \hat{\phi}_{1}\phi_{2}|_{\bx', t'}  \right\rangle_0\;.
\end{align}

\setlength{\tabcolsep}{10pt}
\begin{table*}[t]
	\centering
	\begin{tabular}{@{}c l c c l c@{}}
		\toprule
		& System & Perturbation & Irrelevant if & Conclusion & \\
		\midrule
		\textbf{1.} & \C{4.4cm}{One field (Harris)} & Random $\delta J(\bx)$ & $\nu d>2$ & \C{3.5cm}{Depends} & \cite{Harris1974} \\[0.3cm]
		\textbf{2.} & \C{4.4cm}{Uncoupled identical fields $K_+=K_-=0$, $V_1=V_2$} & $\delta K_-$ & $\gamma<0$ & \C{3.5cm}{Relevant} & \checkmark \\[0.3cm]
		\textbf{3.} & \C{4.4cm}{Uncoupled nonidentical fields $K_+=K_-=0$, $V_1\neq V_2$} & $\delta K_-$ & Always & \C{3.5cm}{Irrelevant} & \\[0.3cm]
		\textbf{4.} & \C{4.4cm}{Reciprocally coupled fields $K_+\neq0$, $K_-=0$, $V_1=V_2$} & $\delta K_-$ & Always & \C{3.5cm}{Irrelevant} & \checkmark \\[0.3cm]
		\textbf{5.} & \C{4.4cm}{Uncoupled identical fields $K_+=K_-=0$, $V_1=V_2$} & Random $\delta K_-(\bx)$ & $\nu d<4\beta$& \C{3.5cm}{Relevantfor 3D Ising, marginal for 2D Ising} & \\[0.3cm]
		\textbf{6.} & \C{4.4cm}{Nonreciprocally coupled fields $K_+=0$, $K_-\neq0$, $V_1=V_2$} & Random $\delta K_-(\bx)$ & $\nu d>2$ & \C{3.5cm}{Irrelevant for 3D swap} & \\
		\bottomrule
	\end{tabular}
	\caption{\textbf{Summary of the results.} The first line is the Harris criterion, which was originally formulated to assess the stability of the ferromagnetic Ising transition with respect to the addition of a local random perturbation in the inter-spin interactions; $J$ refers to nearest-neighbors couplings as in Figure~\ref{fig:CriticalExponents}. Other lines refer to Eq.~\eqref{eq:basic_coupled_equation_1}. A checkmark (\checkmark) indicates results we have numerically tested.}
	\label{figtab:schematic}
\end{table*}

\noindent Since the two fields are independent under the unperturbed dynamics, we can average them separately. Using
\begin{align}
\label{eq:susceptibility_def}
    \frac{\partial\langle\phi_{i}(\bx,t)\rangle_0}{\partial h_i(\bx',t')}\bigg{|}_{h_i=0}=\langle\phi_{i}(\bx,t)\hat{\phi}_{i}(\bx',t') \rangle_0\;,
\end{align} we find
\begin{align}
\label{eq:correction_with_chi}
\delta\langle\phi_1\rangle=-\delta K_- \langle\phi_2 \rangle_{0}\; \chi+O(\delta K_-^2)
\end{align}
where the susceptibility $\chi \equiv \partial \langle \phi_1\rangle/\partial h_1|_{h_1=0}$
is given by the integral of the response function \eqref{eq:susceptibility_def}.

We now compare the scaling of this correction with the unperturbed order parameter when $T\to T_c$ by computing $\delta\langle\phi_1\rangle/\langle\phi_1\rangle_0$.
Because the correction is proportional to $\langle\phi_2 \rangle_0$ and the two fields are identical before the introduction of the perturbation, the correction due to nonreciprocity will dominate (i.e. $\delta\langle\phi_1\rangle/\langle\phi_1\rangle_0 \to \infty$ as $T \to T_c$) as long as the susceptibility diverges at the transition.
In terms of the critical exponent $\gamma$ ($\chi \sim |T-T_c|^{-\gamma}$), our criterion for the relevance of the nonreciprocal perturbation reads $\gamma > 0$ (row 2 in Table~\ref{figtab:schematic}).
This is the case for Model A as well as most physical systems, and in particular for all equilibrium ones \cite{tauber2014critical} \footnote{Equilibrium systems are characterized by a diverging correlation-length $\xi$ in the correlation function. Through the fluctuation-dissipation theorem, this translates in a response function $\chi$ diverging as $\xi^{2-\eta}$ with $\eta>0$. This is not necessarily true out of equilibrium, since FDT need not hold.}.

\begin{figure*}
    \centering
    \includegraphics[width=\linewidth]{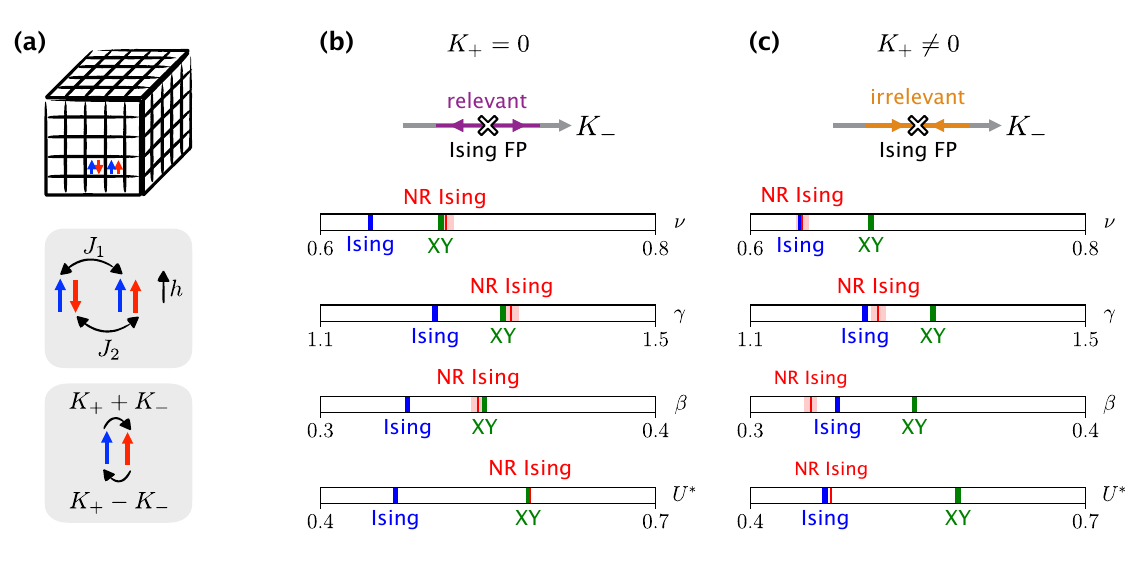}
    \caption{
    \textbf{
    Putting the criterion to the test.}
    (a) An example of a system to which our results apply: two coupled Ising models.
    Note that the couplings $K_\pm$ will be renormalized when going from the discrete model represented here to a field theory.
    (b-c) The relevance of the perturbation, in the RG sense, depends on whether there is a finite reciprocal coupling in the unperturbed system.
    Critical exponents $\nu$, $\gamma$, $\beta$, and Binder cumulant $U^{\star}$ at the critical point for the 3D Nonreciprocal Ising model (red), with $K_+=0$ and $\delta K_-=0.1$ (panel b, corresponding to  row 2 of Table~\ref{figtab:schematic}) taken from Ref.~\cite{Avni2025a} and $K_+=0.5$ and $\delta K_-=0.1$ (panel c, corresponding to  row 4 of Table~\ref{figtab:schematic}), which results from new simulations.
    The exponents are obtained using finite-size scaling, as described in Ref.~\cite{Avni2025a}.
    Standard deviation is represented by a semi-transparent red rectangle. The corresponding values in the 3D Ising model and in the 3D XY model are shown in blue and green respectively, for comparison.
    In the absence of $K_+$, a weak nonreciprocity shifts the critical exponents of the phase transition away from the Ising universality class, whereas when $K_+$ is nonzero, the transition appears to remain within the Ising universality class, in agreement with our analytical results.
    Note that the discrepancy between Ising's and nonreciprocal Ising's $\beta$ in the $K_+ \neq 0$ case is reduced when systematic finite-size errors - omitted in the figure - are taken into account; see SM for details.}
\label{fig:CriticalExponents}
\end{figure*}

This result is in agreement with numerical simulations of two nonreciprocally coupled Ising models \cite{Avni2025}, whose corresponding field theory is similar to the one considered here, with some irrelevant higher order terms.
It is also in agreement with renormalization group studies of similar field theories~\cite{Risler2004,Risler2005,Daviet2024, Zelle2024}.
These works find that the transition to order is destroyed in 2D, while in 3D the ordered phase exhibits persistent oscillations and the critical exponents are significantly modified, becoming compatible with the 3D XY universality class (Fig.~\ref{fig:CriticalExponents}).

\textit{Two critical points ---}
What happens if the two fields have different critical points before the perturbation?
This happens when there are different potentials $V_1$ and $V_2$ on each line in Eq.~\eqref{eq:basic_coupled_equation_1} (or equivalently two different temperatures 
\cite{tauber_effects_2002}).
In an Ising model, this corresponds to different intra-species couplings $J_1$ and $J_2$.
Let us suppose that, when going from the disordered to the ordered phase, $\phi_1$ is the first to encounter the symmetry-breaking transition.
We can carry out the same computation for the correction to $\langle \phi_1 \rangle$, arriving again at Eq. \eqref{eq:correction_with_chi}.
However, around the transition of $\phi_1$, the field $\phi_2$ is still in the disordered phase, so that $\langle \phi_2 \rangle_0=0$, and there is no correction at linear order in $\delta K_-$ (row 3 in Table~\ref{figtab:schematic}).
In general, the coupling to a subcritical field is an irrelevant perturbation -- this also holds in the nonreciprocal case.

\textit{Reciprocally coupled fields ---}
We have so far considered fields that were independent in the absence of the perturbation. What happens when the two fields have a finite symmetric coupling $K_+>0$ in the unperturbed theory?
Two uncoupled Models A have four equivalent minima of the energy: each field can independently have positive or negative magnetization.
The introduction of $K_+$ partially lifts this degeneracy, because states with same-sign magnetizations are now favored.
Upon lowering the temperature the system will select one of these two same-sign states, hence undergoing a phase transition in the Ising universality class, with the order parameter being the sum of the two fields \footnote{Note that considering a reciprocal perturbation $\delta K_+$ would naively lead to the same correction to the order parameter as in Eq.\eqref{eq:nonrec_deviation}. This is however due to a linear shift of the critical point (see SI), which in this case is not prevented by symmetries, and therefore does not imply the relevance of the perturbation.}.
Expressing the perturbed dynamics in terms of the sum and the difference of the two fields, we obtain two field theories with different critical points, coupled only via an irrelevant cubic term (SM).
The nonreciprocal perturbation $\delta K_-$ takes the same antisymmetric form in these new variables.
Hence, as
in the previous section, nonreciprocity is irrelevant for two reciprocally coupled fields (row 4 in Table~\ref{figtab:schematic}).

This prediction is in qualitative agreement with numerical simulations of the nonreciprocal Ising Model performed in \cite{Avni2025}: the addition of a reciprocal coupling between the two species of spins indeed leads to the re-stabilization of the paramagnetic to ferromagnetic transition.
To confirm this, we have measured the critical exponents of this phase transition in $d=3$, and they are compatible with the Ising ones (Figure \ref{fig:CriticalExponents}), see SM for more details. Note that our approach only focuses on the case of small nonreciprocity.
We do expect the critical behavior to change when nonreciprocal interactions become stronger than the reciprocal ones \cite{Avni2025a}.

\textit{Random nonreciprocal perturbations ---}
Our dynamical procedure can also be used to evaluate the relevance of a random perturbation, generalizing the Harris criterion ~\cite{Harris1974} to nonequilibrium settings, building on Ref.~\cite{vojta2016} (see EM and SM).
Going back to nonreciprocally coupled fields, we first consider a space-dependent random antisymmetric perturbation $\delta K_-(\bx)$, coupling $\phi_1(\bx)$ and $\phi_2(\bx)$.
Since $\delta K_-(\bx)$ is random and symmetric around zero, it can't shift the critical point at first order.
Here, $\delta K_-(\bx)$ is Gaussian distributed with mean zero and delta correlations $\overline{\delta K_-(\bx)\delta K_-(\bx')}=\delta \sigma_K^2 \delta(\bx-\bx')$, with the overbar indicating an average over the quenched disorder.
We then compute the correction to the order parameter $\langle \phi_i \rangle$.
Since the perturbation is random, so is the correction.
Because it averages to zero ($\overline{\delta\langle\phi_i(\bx,t)\rangle}=0$), its typical size is characterized by its variance
(see SM)
\begin{align*}
        \frac{\overline{\delta \langle \phi_i(\bx,t)\rangle^2 }}{\langle \phi_{i}\rangle^2_0}
        \sim \delta \sigma_K^2 \int\displaylimits_{\bx',t',t''} \left\langle \frac{\partial \phi_i(\bx,t)}{\partial h_i(\bx', t')} \right\rangle_0 \left\langle \frac{\partial \phi_i(\bx,t)}{\partial h_i(\bx', t'')} \right\rangle_0.
    \end{align*}

Close to criticality, the response function scales as \cite{cardy1996}
\begin{align}
\label{eq:charac}
        \left\langle\frac{\partial \phi_i(\bx,t)}{\partial h_i(\bx', t')}\right\rangle_0=\xi^{-d-z+\frac{\gamma}{\nu} } f\left(\frac{\bx-\bx'}{\xi}, \frac{t-t'}{\tau}\right)
    \end{align}
where $d$ is the spatial dimension while $\xi$ and $\tau$ are respectively the correlation length and time, diverging at the transition as $\xi\sim|T-T_c|^{-\nu}$ and $\tau\sim \xi^z$.
In addition, the order parameter scales as $\langle \phi\rangle_0\sim|T-T_c|^\beta$ in which $\beta$ is the associated critical exponent, related to the others by the identity $\gamma=d\nu-2\beta$ obtained from Widom's and Fisher's scaling relations.
Putting these together, we find that
\begin{align}
        \frac{\overline{\delta \langle \phi_i(\bx,t)\rangle^2 }}{\langle \phi_{i}\rangle^2_0}
        \sim
        \delta \sigma_K^2 \, (T-T_c)^{4\beta-\nu d}
\end{align}
Hence, the introduction of random nonreciprocal interactions is relevant if
\begin{align}
    2\beta - \frac{\nu d}{2}<0.
\end{align}
This inequality is satisfied by the 3D Ising model, so we expect its universality class to change in the presence of random nonreciprocity.
For the 2D Ising model, the left hand side of the inequality is exactly zero, so that the correction is marginal: a more refined analysis is required, similarly to the Harris criterion for the 2D random bond Ising model \cite{Shalaev1994}.
This is summarized in row 5 of Table~\ref{figtab:schematic}.

\textit{Perturbing a nonequilibrium phase transition ---}
Lastly, we illustrate that our method encompasses situations where the unperturbed phase transition is already out of equilibrium.
To do so, we take as the unperturbed system Eq.~\eqref{eq:basic_coupled_equation_1} with a finite nonreciprocal coupling $K_-$, and add as a perturbation a random inhomogeneous $ \delta K_-(\bx)$, where $\delta K_-(\bx)$ is a delta-correlated Gaussian variable.
We have previously shown that a small nonreciprocal coupling $K_-$ changes the critical behavior of the paramagnet/ferromagnet transition present when $K_-=0$.
The resulting phase transition at finite $K_{-}$ cannot be predicted from our criterion, but has been studied through renormalization group calculations \cite{Risler2004,Risler2005,Daviet2024,Zelle2024} and Monte-Carlo simulations \cite{Avni2025,Avni2025a}.
This phase transition, which is believed to fall in the $XY$ universality class, separates a disordered (paramagnet) phase from a time-dependent oscillating phase, dubbed \textit{swap phase}, where the fields $\phi_1$ and $\phi_2$ homogeneously and coherently oscillate in time \cite{Avni2025,Avni2025a}.

Our analysis (see SM)  shows that $\delta K_-(\bx)$ is relevant whenever $\frac{\nu d}{2} < 1$.
Thus we recover the Harris criterion’s form, but with the crucial distinction that $\nu$ now refers to the critical exponent of the unperturbed nonequilibrium dynamical transition.
The unperturbed system is believed to fall in the XY universality class \cite{Avni2025a} in $d=3$, leading to $\nu= 0.672$~\cite{Pelissetto2002,Campostrini2001}. Therefore, the perturbation is irrelevant (row 6 in Table~\ref{figtab:schematic}).

\textit{Conclusion ---}
To sum up, we have derived perturbative criteria, $\grave{a}$ la Harris, to assess whether nonreciprocal perturbations are relevant in a critical system.
In the case of two-species nonreciprocity, which we have focused on, symmetries greatly simplifies the analysis and makes obtaining the scaling of the correction $\delta \langle \mathcal{O} \rangle$ in Fig.~\ref{fig:pipeline} analytically tractable.
This generalizes to other (equilibrium or nonequilibrium) systems symmetric under exchange of the fields and inversion symmetry that exhibit a continuous phase transition characterized by a local order parameter (EM).
In the case of $O(N)$ models, for instance, we recover the results of renormalization group calculations \cite{Daviet2024, Zelle2024} indicating that the equilibrium critical point is destabilized by an infinitesimal nonreciprocal coupling.

The general procedure presented in Figure \ref{fig:pipeline} can be applied to other unperturbed critical system and perturbations, but obtaining the scaling of the correction could be nontrivial.
In order to handle dynamical properties (for instance dynamical critical exponents) or conserved order parameter fields, the procedure would also have to be tweaked to replace the order parameter $\mathcal{O}$ with a correlation function.
Finally, we emphasize that the criteria discussed in this work do not assess any change (e.g. new phases or phase transitions) away from criticality and beyond perturbative corrections.

\medskip

\begin{acknowledgments}
This work was supported by the Simons Foundation Grant No. 454935 (G.B.).
V.V. acknowledges partial support from the Army Research Office under grant  W911NF-22-2-0109 and W911NF-23-1-0212 and the Chan Zuckerberg Initiative.
M.F. and V.V acknowledge partial support from the France Chicago center through a FACCTS grant.
This research was partly supported from the National Science Foundation through the Center for Living Systems (grant no. 2317138), and by grants from the NSF (DMS-2235451) and Simons Foundation (MP-TMPS-00005320) to the NSF-Simons National Institute for Theory and Mathematics in Biology (NITMB). Computing resources were provided by the University of Chicago Research Computing Center.
\end{acknowledgments}

\bibliography{refs}

\newpage
\appendix

\onecolumngrid
\vspace{5mm}
\begin{center}
    \large \textbf{End Matter}
\end{center}
\twocolumngrid

\section{Nonreciprocal perturbations nonparallel to the critical line}
\begin{figure}[h]
    \centering
    \includegraphics{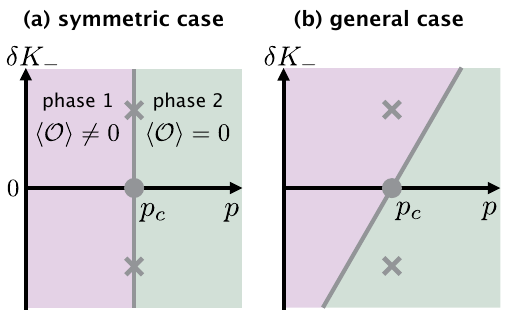}
    \caption{
    Panel (a) shows a symmetric case, like the ones studied in the main text, in which switching on the nonreciprocal perturbation moves the system parallel to the critical line. Panel (b) shows the general case in which the shift is not parallel, and hence move the system closer or away from criticality. This leads to a trivial singular contribution to the order parameter that has to be subtracted to assess the relevance of the perturbation.
    Symmetries of the stochastic action can enforce the symmetric case, at least to first order in $\delta K_{-}$.
    In these cases, a constraint of the form $|\langle \mathcal{O}\rangle_{\delta K_{-}}| = |\langle \mathcal{O}\rangle_{-\delta K_{-}}|$ where $\mathcal{O}$ is the order parameter field can be obtained from the symmetries of the action and the order parameter field.
    This constraint is not compatible with case (b) because the values of the order parameters at the points marked by crosses are not compatible with the constraint.
    In the cases considered in the main text, $p$ corresponds to the temperature $T$.
    }
    \label{fig:parallelPerturbation}
\end{figure}
The procedure described in the main text aims at investigating whether adding nonreciprocity is a singular perturbation which alters the critical behavior. In the absence of such perturbation, $\langle \phi \rangle \sim \epsilon ^\beta $, where $\epsilon=T-T_c$ is the distance from the critical point in absence of the perturbation and $\beta$ the associated critical exponent.
In the presence of even infinitesimal nonreciprocity, the critical behavior becomes $\langle \phi \rangle \sim \epsilon ^{\beta'} $, with possibly $\beta'\ne \beta$.
It is important, while doing the comparison between perturbed and unperturbed scaling, to work at {\it fixed} $\epsilon$.
In the main text, we examine systems with nonreciprocal perturbations for which the critical temperature remains unchanged to linear order in $\delta K_-$, while the critical point is approached along a direction perpendicular to the critical line (see Fig.~\ref{fig:parallelPerturbation}a).
This is guaranteed by symmetry considerations (see next section).
Consequently, we indeed perform the comparison at fixed $\epsilon$.
In more general cases (for example if we add a small reciprocal interaction between the two fields, see SI and footnote [56]), like Fig.~\ref{fig:parallelPerturbation}b, the nonreciprocal perturbation is not parallel to the critical line, and hence does not keep $\epsilon$ fixed.
In order to take into account the effect of the perturbation on the order parameter at fixed $\epsilon$ one has thus to focus on:
\[
\left.\frac{d\langle\phi \rangle}{dK_-}\right|_{\epsilon}=\frac{d\langle\phi \rangle}{d T}\frac{dT}{dK_-}+\frac{d\langle\phi \rangle}{d K_-}
\] where $\frac{dT}{dK_-}$ is chosen to keep $\epsilon$ constant. Since $\epsilon=T-T_c(K_-)$, this leads to $\frac{dT}{dK_-}=\frac{dT_c}{dK_-}$. Therefore, our procedure can be generalized to the cases illustrated in the lower panel of Fig. 3 by comparing the unperturbed critical behavior to the term:
\begin{equation}
    \left(\left.\frac{d\langle\phi \rangle}{d T}\right|_{\delta K_-=0}\left.\frac{dT_c}{dK_-}\right|_{\delta K_-=0}+\left.\frac{d\langle\phi \rangle}{d K_{-}}\right|_{\delta K_-=0}\right)\delta K_-
    \label{eq:corr}
\end{equation}
The cases analyzed in the main text, in which an underlying symmetry guarantees that $T_c$ is not shifted at linear order \footnote{It's instructive to see the effect of the extra term in Eq.~\eqref{eq:corr} when $T_c$ is shifted at linear order but the critical behavior in not altered: when computing the perturbative corrections to $\langle \phi \rangle \sim (T-T_c)^\beta$ this leads to a contribution $(\beta-1)(T-T_c)^{\beta-1}\left.\frac{dT_c}{dK_-}\right|_{\delta K_-=0} \delta K_-$. This term is more singular than the nonperturbed one, but it is not related to a change in the critical properties. The first term in Eq.~\eqref{eq:corr} correctly cancels it out.}, correspond to $\left.\frac{dT_c}{dK_-}\right|_{\delta K_-=0}=0$.
All terms in Eq.\eqref{eq:corr} can be obtained from correlation and response functions of the {\it unperturbed} critical point \cite{zinn2021quantum}. Therefore the relevance of the nonreciprocal perturbation can be assessed only using the critical exponents of the unperturbed system.

\section{Symmetry requirements and general setting}\label{sec:symmetry}

The analysis developed in the main text can be applied to any pair of identical fields undergoing a second order phase transition whose dynamics is symmetric under field exchange and that are only coupled by an antisymmetric perturbation.
In the following we provide more details on these requirements.

\textit{Symmetries of the action ---}
Let us consider a system whose dynamics can be described by an action $\cS_0(\phi, \hat{\phi})$, where $\phi$ is not necessarily scalar.
We assume the action to be invariant under a symmetry group $\cG$, so that $\cS_0(\cR(\phi,\hat{\phi}))=\cS_0(\phi, \hat{\phi})$ for all $\cR\in \cG$.
We assume that $\cG$ contains  the inversion operation $(\phi,\hat{\phi}) \to (-\phi,-\hat{\phi})$. Examples include the $O(N)$ symmetry, but also the dihedral group $D_{2n}$, i.e. the symmetry group of a regular polygon with an even number of sides.
We suppose that the system undergoes a second order phase transition, after which the considered symmetry is spontaneously broken.

We then study two identical copies of the system, $\phi_1$ and $\phi_2$.
If the fields are not coupled, their dynamics is described by the sum of the two actions,
\begin{align}
	\cS_0(\phi_1, \hat{\phi}_1,\phi_2, \hat{\phi}_2)=\cS_0(\phi_1, \hat{\phi}_1)+\cS_0(\phi_2, \hat{\phi}_2)\ .
\end{align}
This action is invariant under the exchange of the two fields
\begin{align}
	\cS_0(\phi_1, \hat{\phi}_1,\phi_2, \hat{\phi}_2)=\cS_0(\phi_2, \hat{\phi}_2,\phi_1, \hat{\phi}_1)\;,
\end{align}
as well as under the independent transformation of each of the two fields: $\forall \cR_1, \cR_2\in \cG$,
\begin{align}
	\cS_0(\cR_1(\phi_1,\hat{\phi}_1),\cR_2(\phi_2,\hat{\phi}_2))=\cS_0(\phi_1, \hat{\phi}_1,\phi_2, \hat{\phi}_2)
\end{align}

We now consider a perturbed system with action
\begin{equation}
    \cS(\phi_1, \hat{\phi}_1,\phi_2, \hat{\phi}_2)=\cS_0
    +
    \epsilon\, \delta \cS
\end{equation}
In examples 2--4 of Table \ref{figtab:schematic},
$\epsilon = \delta K_{-}$.
We require that the perturbation $\delta \cS$  respects the following conditions:
\begin{enumerate}
	\item After the perturbation, the system is still invariant under a \textit{simultaneous} transformation of the two fields. This can be imposed by requiring that $\forall \cR\in \cG$, $\delta \cS(\cR(\phi_1,\hat{\phi}_1),\cR(\phi_2,\hat{\phi}_2))=\delta \cS(\phi_1, \hat{\phi}_1,\phi_2, \hat{\phi}_2)$.
	\item Exchanging the two fields changes the sign of the perturbation, $\delta \cS(\phi_2, \hat{\phi}_2,\phi_1, \hat{\phi}_1)=- \delta \cS(\phi_1, \hat{\phi}_1,\phi_2, \hat{\phi}_2)$.
    \item Inverting one of the two fields changes the sign of the perturbation, $\delta \cS(\phi_1, \hat{\phi}_1,-\phi_2, -\hat{\phi}_2)=- \delta \cS(\phi_1, \hat{\phi}_1,\phi_2, \hat{\phi}_2)$.
    Combining this property with the previous one implies that exchanging the two fields and inverting one of the two leaves the perturbation unchanged, $\delta \cS(\phi_2, \hat{\phi}_2,-\phi_1, -\hat{\phi}_1)= \delta \cS(\phi_1, \hat{\phi}_1,\phi_2, \hat{\phi}_2)$.
    Since such a transformation leaves also the unperturbed action $\cS_0$ unchanged, it is a symmetry of the \textit{perturbed} system.
	\item $\delta \cS$ is an analytic function of its arguments
\end{enumerate}

The symmetries of the perturbed system include (i) thanks to condition 1, the same symmetry as one of the original two fields, and (ii) thanks to conditions 2-3, the symmetry $(\phi_1, \phi_2)\to (\phi_2, -\phi_1)$.
The breaking of the first symmetry can therefore be detected by studying the average value of \textit{either} of the two fields, which are order parameter fields of the unperturbed phase transition.
In addition, the second symmetry prevents the critical point from being shifted at linear order in the perturbation.
Indeed, a reversal of $\delta \cS$ ($\delta \cS \to - \delta \cS$) amounts to exchanging the labels of the two fields (which are interchangeable thanks to the symmetry $(\phi_1, \phi_2)\to (\phi_2, -\phi_1)$), hence leaving the critical point unchanged: its shift must therefore be even in the parameter defining the perturbation.
This arises from the constraint $|\langle \mathcal{O}\rangle_{\epsilon}| = |\langle \mathcal{O}\rangle_{-\epsilon}|$ where $\mathcal{O}$ is the order parameter ($\phi_1$ or $\phi_2$), which is a direct consequence of the symmetry of $\mathcal{S}$.
Since we assumed analyticity (condition 4), the shift must be at least of quadratic order.

Conditions 1 to 4 ensure two key properties: (1) the average value of the fields is the order parameter, (2) a small nonreciprocal perturbation shifts the system parallel to the critical line.
Since our results derived in the main text relied on these two properties,
they equally hold for any system complying with conditions 1 to 4.
Hence, the introduction of nonreciprocity is relevant whenever the susceptibility diverges at the phase transition as discussed in a concrete case below.

\textit{A concrete example ---}
To perform the computation, we need to be more specific about the functional form of the perturbation.
In systems with $O(N)$ symmetry, a particular form that the perturbation can take is
\begin{align*}
	\delta \cS = \delta K_-\left(\hat\phi_1 \phi_2 -\hat \phi_2 \phi_1 \right)f\left(\phi_1^2, \phi_2^2, \left(\hat\phi_1 \phi_2 -\hat \phi_2 \phi_1\right)^2\right)\ ,
\end{align*}
where $f$ is any analytic function symmetric under the exchange of the first two arguments.
For such a $\delta \cS$, we can use scaling arguments to show that the dominant contribution remains the same as in the main text.
Let us consider $f=\phi_1^2+\phi_2^2$ as an example.
It corresponds to the following perturbed dynamics
\begin{align*}
	\begin{split}
		\partial_t\phi_1&=\nabla^2\phi_1 - V^{\prime}(\phi_1) + \eta_1  +h_1+ \delta K_- \phi_2\left(\phi_1^2+\phi_2^2\right)\;,\\
		\partial_t\phi_2&=\nabla^2\phi_2 - V^{\prime}(\phi_2) + \eta_2 +h_2- \delta K_- \phi_1 \left(\phi_1^2+\phi_2^2\right)\;.
	\end{split}
\end{align*}
The correction to the order parameter is then given by
\begin{align}
	&\delta\langle\phi_1(\bx,t)\rangle =\\&- \delta K_- \left\langle \phi_1(\bx,t) \int_{\bx', t'} \left(\hat{\phi}_1\phi_2-\hat{\phi}_2\phi_1\right)\left(\phi_1^2+\phi_2^2\right) \bigg|_{\bx', t'} \right\rangle_0 \nonumber
\end{align}
Since the unperturbed fields are uncoupled, averages can be computed independently for $\phi_1$ and $\phi_2$.
Note that terms that contain $\hat \phi_2$ do not contribute, because $\hat \phi_2(x,t)$ corresponds to the response to a perturbation at a time infinitesimally successive to $t$, so that $\langle \hat{\phi}_2\phi_2^2 \big|_{\bx', t'} \rangle_0=0$.
We also remark that $\phi_1^2$ and $\phi_2^2$  take a finite value at the transition, therefore they do not change the critical behavior of the dominant term (see SM).
The perturbation is thus relevant whenever the susceptibility diverges.

\section{Nonequilibrium Harris criterion}
\label{sec:spirit}

We now sketch the derivation of the Harris criterion in a dynamical formulation, which allows us to generalize it beyond equilibrium systems.
Our approach is similar to the one used in \cite{vojta2016} for generalizing the Harris criterion to arbitrary spatio-temporal disorder. It is also a generalization of \cite{sarlat2009}, which developed a dynamical formulation for the standard Harris case.

We consider a field theory perturbed by a random variation of the mass term $ \delta m(\bx)$
\begin{align}
\label{eq:evolution1field}
\partial_t\phi&=\nabla^2\phi - V^{\prime}(\phi) + \eta +\delta m(\bx)\phi \;,
\end{align}
where $V(\phi)$ is a potential function and $\eta$ is a Gaussian white noise of amplitude $T$.
The Gaussian disorder $\delta m(\bx)$ is $\delta$-correlated in space
\begin{align}
    \overline{\delta m(\bx)\delta m(\bx')}=\delta \sigma_m^2\delta(\bx-\bx')\;,
\end{align}
where the overline indicates averaging over quenched disorder.
Using the MSRDJ approach, the probability of observing a given configuration of the field can be expressed as
\begin{align}
\mathcal{P}\left(\{\phi\}\right)=\int D[\hat{\phi}]e^{-\int d\bx dt \mathcal{S}}\;,
\end{align}
where $\hat{\phi}$ is an auxiliary field. The action $\cS$ can be decomposed as $\cS=\cS_0[\hat{\phi},\phi] + \delta \cS[\hat{\phi},\phi]$, with $\cS_0$ being the action in the absence of perturbations and $\delta \cS$ containing the perturbative terms
\begin{align}
\label{eq:action_0}
    \cS_0=&\hat{\phi}\left(\partial_t \phi-\bnab^2\phi+V^{\prime}(\phi)\right) + \frac{T}{2}\hat{\phi}^2 \;,\\
\label{eq:action_pert}
    \delta \cS =& \delta m (\bx)\hat{\phi}\phi \;.
\end{align}
To see whether the perturbation modifies the critical properties of the system, we compute the correction to a generic observable $\cO$.
This observable can correspond, for instance, to the magnetization $\phi(\bx, t)$.
Expanding the exponential of the action to first order in $\sigma_m$, we find
\begin{align}
\delta \langle\cO(\bx,t)\rangle = - \langle \cO(\bx,t) \int \delta \cS|_{\bx', t'} d\bx' dt'  \rangle_0 \;.
\end{align}
This correction can be further expressed as
\begin{align}
\delta\langle\cO(\bx,t)\rangle =- \langle \int \delta m (\bx')G(\bx-\bx') d\bx'   \rangle_0 \; ,
\end{align}
where $G(\bx-\bx')=\langle \frac{\delta \cO(\bx)}{\delta m(x')}\rangle_0$ is the response function of $\cO $ with respect to a local variation of the linear term.
Noting that the first order correction averages to 0, we set up to obtain its typical amplitude by deriving its variance as
\begin{align}
\label{eq:correctionvariance1}
\overline{\delta\langle\cO\rangle^2} =\delta\sigma_m^2 \int G(\bx')^2 d\bx'  \; .
\end{align}
Using critical scaling properties, we can show that \cite{cardy1996}
\begin{align}
    \label{eq:response_ising2}
    G(\bx')\sim \xi^{-(d+(\beta-1)/\nu)}f\left(\frac{\bx'}{\xi}\right)\;.
\end{align}
Inserting this expression in \eqref{eq:correctionvariance1} we obtain the typical amplitude of the correction as
\begin{align}
\label{eq:typical_size_noneq}
\sqrt{\overline{\delta\langle\cO\rangle^2}} \sim  \delta\sigma_m (T-T_c)^{d\nu/2+\beta-1}\;.
\end{align}
Comparing \eqref{eq:typical_size_noneq} to the behavior of $\langle \cO\rangle$ in the unperturbed system, we deduce the correction to dominate when $d\nu/2<1$: this is the Harris criterion.

As in the case of nonreciprocal perturbations, note that symmetry prevents any shift of the critical temperature to linear order.
Indeed, since the distribution of $\delta m$ remains symmetric around 0, reversing the sign of the perturbation leaves the system unchanged.

\onecolumngrid
\vspace{5mm}
\begin{center}
    \large \textbf{Supplemental material}
\end{center}

\makeatletter
\def\c@secnumdepth{0}
\def\thesection{\arabic{section}}
\def\thesubsection{\arabic{section}.\arabic{subsection}}
\let\@sectioncntformat\relax
\let\@hangfrom@section\relax

In this supplemental material, we report some additional details on the computations performed in the main text.
In Section \ref{app:ft_const_nonrec_coupl}, we derive the field theory that we considered in this work, and the most general form of the correction to the order parameter.
In Section \ref{app_uncoupled}, we focus on the case in which two identical uncoupled fields are perturbed by antisymmetric interactions.
In Section \ref{app_constant_nr_unperturbed_r}, we add a constant nonreciprocal coupling.
In Section \ref{random_nr} and \ref{app:nonrec_with_quenched}, we consider random nonreciprocal perturbations.
In Section \ref{mcmc} we give details on the calculation of the critical exponents from Monte-Carlo simulations.
In Section \ref{example_nonrelaxational} we show how our treatment can be applied when the unperturbed critical system follows nonrelaxational critical dynamics.
In Section \ref{app:DP} we illustrate the role of inversion symmetry.

\section{Field theory derivation}
\label{app:ft_const_nonrec_coupl}
In this appendix, we detail the derivation of the relevance criterion for field theory (1) with constant nonreciprocal couplings.
We start by deriving the generic action valid for all types of couplings studied in this paper, namely for
\begin{align}
    K_{12}=K_++(K_-+\delta K_-(\bx))\;, && K_{21}=K_+-(K_-+\delta K_-(\bx))\;.
\end{align}
The case of constant nonreciprocal couplings studied in the main text thus corresponds to $K_+=K_-=0$ and $\delta K_-(\bx)=\delta K_-$, independently of the position.
We start by recalling the time evolution of the fields
\begin{align}
\begin{cases}
    \partial_t\phi_1=\nabla^2\phi_1 - V^{\prime}(\phi_1) + \eta_1 + K_{12} \phi_2 + h_1\;, \\
    \partial_t\phi_2=\nabla^2\phi_2 - V^{\prime}(\phi_2) + \eta_2 + K_{21} \phi_1 + h_2\;,
\end{cases}
\label{eq:dynevSM}
\end{align}
We first write the probability $\mathcal{P}\left(\{\phi_1,\phi_2\}\right)$ of observing a trajectory of the fields $\{\phi_1,\phi_2\}$ by using the MSRJD formalism. It reads
\begin{align}
    \mathcal{P}\left(\{\phi_1,\phi_2\}\right)=&\langle \delta\left(\partial_t\phi_1 -\nabla^2\phi_1 + V^{\prime}(\phi_1) - \eta_1 +K_{12}\phi_2+ h_1\right)\delta\left(\partial_t\phi_2 -\nabla^2\phi_2 + V^{\prime}(\phi_2) - \eta_2 +K_{21}\phi_1 + h_2\right) \rangle\;,
\end{align}
where the mean value $\langle\cdot\rangle$ runs over all possible trajectories of $\eta_1$ and $\eta_2$ while $\delta (\cdot)$ represents the Dirac delta.
Using the integral representation of the Dirac deltas allows us to introduce the imaginary auxiliary fields $\hat{\phi}_1$ and $\hat{\phi}_2$ as
\begin{align}
    \label{eq:path_1}
    \mathcal{P}\left(\{\phi_1,\phi_2\}\right)=\int D[\hat{\phi}_1,\hat{\phi}_2,\eta_1,\eta_2] \exp\left(-{\int\mathcal{S}_{\eta}(\phi_1,\phi_2,\hat{\phi}_1,\hat{\phi}_2)d\bx dt}\right)\exp\left(-\frac{1}{2T}\int \eta_1^2 d\bx dt-\frac{1}{2T}\int \eta_2^2 d\bx dt\right)\;,
\end{align}
where the action $S_{\eta}$ reads
\begin{align}
     \mathcal{S}_{\eta}=\hat{\phi}_1\left(\partial_t \phi_1-\nabla^2\phi_1+V^{\prime}(\phi_1)+K_{12}\phi_2-\eta_1 + h_1\right)+\hat{\phi_2}\left(\partial_t \phi_2-\nabla^2\phi_2+V^{\prime}(\phi_2)+K_{21} \phi_1-\eta_2+h_2\right)\;.
\end{align}
As $\cS_{\eta}$ is linear in the noises $\eta_j$'s, we can use the Gaussian integration formula $\int_{-\infty}^{\infty}e^{-ax^2-bx}dx=\sqrt{\frac{\pi}{a}}e^{\frac{b^2}{4a}}$ with $b=-\hat{\phi}_j$ to perform the integration over the $\eta_j$'s in \eqref{eq:path_1}. We obtain
\begin{align}
    \label{eq:path_2}
    \mathcal{P}\left(\{\phi_1,\phi_2\}\right)=\int D[\hat{\phi}_1,\hat{\phi}_2]\exp\left(-\int\mathcal{S}(\phi_1,\phi_2,\hat{\phi}_1,\hat{\phi}_2)d\bx dt\right)\;,
\end{align}
where $\mathcal{S}$ is given by
\begin{align}
\label{eq:generic_action_undevelopped}
\mathcal{S}(\hat{\phi}_1,\hat{\phi}_2,\phi_1,\phi_2)=\cS_{0}(\hat{\phi}_1,\hat{\phi}_2,\phi_1,\phi_2) + \delta \cS(\hat{\phi}_1,\hat{\phi}_2,\phi_1,\phi_2)\;,
\end{align}
with $\cS_0(\hat{\phi}_1,\hat{\phi}_2,\phi_1,\phi_2)$ and $\delta \cS$ reading
\begin{align}
\cS_0(\hat{\phi}_1,\hat{\phi}_2,\phi_1,\phi_2)=&\cS_0(\hat{\phi}_1,\phi_1,h_1) + \cS_0(\hat{\phi}_2,\phi_2,h_2) + (K_++K_-) \hat{\phi}_1\phi_2+ (K_+-K_-) \hat{\phi}_2\phi_1\;,  \\
\delta \cS(\hat{\phi}_1,\hat{\phi}_2,\phi_1,\phi_2)=&\delta K_-(\bx)\left(\hat{\phi}_1\phi_2-\hat{\phi}_2\phi_1\right)\;,
\end{align}
and $\cS_0(\hat{\phi},\phi)$ is the action of an uncoupled field given by
\begin{align}
    \cS_0(\hat{\phi},\phi,h)=&\hat{\phi}\left(\partial_t \phi-\bnab^2\phi+V^{\prime}(\phi)\right) - \frac{T}{2}\hat{\phi}^2 + \hat{\phi}h\;. \label{eq:unperturbedsinglefieldaction}
\end{align}
The auxiliary fields $\hat{\phi}_i(\bx, t)$ are also called ``response fields'' because they generate response functions according to
\begin{align}
    \langle \phi_j(\bx, t)\hat \phi_j(\bx', t')\rangle= \frac{\delta \langle \phi_j(\bx, t)\rangle}{\delta h_j(\bx', t')}\Bigg|_{h_j=0}\ .
\end{align}
We compute the average magnetization as
\begin{align}
\nonumber
\langle\phi_{j}\left(\bx,t\right)\rangle & =\int D\left[\hat{\phi}_1,\hat{\phi}_2,\phi_1,\phi_2\right]\phi_{j}\left(\bx,t\right)e^{-\int \left(\cS_{0}(\hat{\phi}_1,\hat{\phi}_2,\phi_1,\phi_2)+\delta \cS\right) d\bx' dt'}\\
\nonumber
 & \approx\int D\left[\hat{\phi}_1,\hat{\phi}_2,\phi_1,\phi_2\right]\phi_{j}\left(\bx,t\right)e^{-\int \cS_{0}(\hat{\phi}_1,\hat{\phi}_2,\phi_1,\phi_2)d\bx' dt'}\left(1-\int \delta \cS d\bx' dt'+\mathcal{O}\left(\delta K_-^2\right)\right)
 \\
 \label{eq:kubo-decomposition}
 & \approx\langle\phi_{j}\left(\bx,t\right)\rangle_{0}-\langle\phi_{j}\left(\bx,t\right)\int\delta \cS d\bx' dt'\rangle_{0}+\cO(\delta K_-^2)\;,
\end{align}
where $\langle \cdot \rangle_0$ implies averaging over the $\cS_0$ action only.
We finally obtain
\begin{align}
\label{eq:perturbation_generic}
    \delta \langle \phi_{j}\left(\bx,t\right) \rangle = \langle \phi_{j}\left(\bx,t\right) \rangle -  \langle \phi_{j}\left(\bx,t\right) \rangle_0 = -\int \langle \phi_{j}\left(\bx,t\right) \delta K_-(\bx)\left( \hat{\phi}_1\phi_2 - \hat{\phi}_2\phi_1\right) \Big|_{\bx', t'}\rangle_0 d\bx' dt' +\cO(\delta K_-^2) \;.
\end{align}

\section{Uncoupled fields}
\label{app_uncoupled}
In this part, we assume that $\phi_1$ and $\phi_2$ are uncoupled before the introduction of the nonreciprocity, i.e. $K_+=K_-=0$.
This implies that averages over the unperturbed action $\cS_0$ can be performed independently for the two fields.
The integrand of \eqref{eq:perturbation_generic} can then be evaluated as
\begin{align}
\begin{split}
        \langle \phi_{1}\left(\bx,t\right) \delta \cS\rangle_0=& \delta K_-(\bx')\left(\langle \phi_{1}\left(\bx,t\right) \hat{\phi}_1(\bx',t')\rangle_0\langle \phi_2(\bx',t')\rangle_0 - \langle \hat{\phi}_2(\bx',t')\rangle_0 \langle \phi_{1}\left(\bx,t\right) \phi_1(\bx',t')\rangle_0\right) \\
    \label{eq:result_integrand_const_nonrec}
    =& \delta K_-(\bx')\langle \phi_{1}\left(\bx,t\right) \hat{\phi}_1(\bx',t')\rangle_0\langle \phi_2(\bx',t')\rangle_0\;.
\end{split}
\end{align}
To obtain \eqref{eq:result_integrand_const_nonrec}, we have further remarked that $\langle \hat{\phi}_i \rangle_0=0$ since
\begin{align}
    \langle \hat{\phi}_i \rangle_0 = \frac{\delta}{\delta h_i(\bx')}\bigg|_{h_i=0} \int D\left[\hat{\phi}_1,\hat{\phi}_2,\phi_1,\phi_2\right]e^{-\int S_{0}(\hat{\phi}_1,\hat{\phi}_2,\phi_1,\phi_2) d\bx dt}=\frac{\delta}{\delta h_i(\bx')}\bigg|_{h_i=0} 1 = 0\;.
\end{align}
Note that, while we considered the case $j=1$ in \eqref{eq:result_integrand_const_nonrec}, the same derivation can be straightforwardly extended to $j=2$.
Generalizing to $j=1,2$ and replacing the response field with the corresponding derivative in $h_j$, we obtain
\begin{align}
    \langle \phi_{j}\left(\bx,t\right) \delta \cS\rangle_0=& \epsilon_{ji}\delta K_-(\bx')\frac{\delta \langle \phi_j (\bx,t)\rangle_0}{\delta h_j(\bx',t')}\bigg|_{h_j=0}\langle \phi_i(\bx',t')\rangle_0\;.
\end{align}
The correction at first order in $\delta \cS$ thus reads
\begin{align}
\label{eq:perturbation_indep}
    \delta \langle \phi_{j}\left(\bx,t\right) \rangle = -\epsilon_{ji} \int \delta K_-(\bx')\frac{\delta \langle \phi_j (\bx,t)\rangle_0}{\delta h_j(\bx',t')}\bigg|_{h_j=0}\langle \phi_i(\bx',t')\rangle_0 d\bx' dt' \;,
\end{align}
The unperturbed system is translationally invariant in time and space, therefore $\langle \phi_i(\bx',t')\rangle_0$ is constant and can be pulled out of the integral.
If $\delta K_-$ is also uniform in space, the integration only concerns the response function and we obtain
\begin{align}
\label{eq:perturbation_indep_uni}
    \delta \langle \phi_{j}\left(\bx,t\right) \rangle = -\epsilon_{ji}\delta K_-\langle \phi_i\rangle_0 \int \frac{\delta \langle \phi_j (\bx,t)\rangle_0}{\delta h_j(\bx',t')}\bigg|_{h_j=0} d\bx' dt'= -\epsilon_{ji}\delta K_-\langle \phi_i\rangle_0  \frac{\delta \langle \phi_j \rangle_0}{\delta h_j}\bigg|_{h_j=0}= -\epsilon_{ji}\delta K_-\langle \phi_i\rangle_0  \chi \;,
\end{align}

\subsection*{Alternative form of the perturbation}

To show that the argument holds for generic forms of the perturbation, in the End Matter we also considered
\begin{align*}
	\delta \cS = \delta K_-\left(\hat\phi_1 \phi_2 -\hat \phi_2 \phi_1 \right)\left(\phi_1^2 + \phi_2^2\right)\ .
\end{align*}
The first order correction to the order parameter becomes
\begin{align}
	\delta\langle\phi_1(\bx,t)\rangle =- \delta K_- \left\langle \phi_1(\bx,t) \int_{\bx', t'} \left(\hat{\phi}_1\phi_2-\hat{\phi}_2\phi_1\right)\left(\phi_1^2+\phi_2^2\right) \bigg|_{\bx', t'} \right\rangle_0 \nonumber
\end{align}
Since the unperturbed fields are uncoupled, averages can be computed independently for $\phi_1$ and $\phi_2$.
Note that terms that contain $\hat \phi_2$ do not contribute, because $\hat \phi_2(x,t)$ corresponds to the response to a perturbation at a time infinitesimally successive to $t$, so that $\langle \hat{\phi}_2\phi_2^2 \big|_{\bx', t'} \rangle_0=0$ due to causality.
The only terms that contribute are therefore
\begin{align}
	\delta\langle\phi_1(\bx,t)\rangle =- \delta K_-\int_{\bx', t'}\left( \left\langle \phi_1(\bx,t)\hat{\phi}_1(\bx', t')\phi_1^2(\bx', t') \right\rangle_0  \left\langle \phi_2(\bx', t') \right\rangle_0 + \left\langle \phi_1(\bx,t)\hat{\phi}_1(\bx', t') \right\rangle_0  \left\langle \phi_2^3(\bx', t') \right\rangle_0\right) .\nonumber
\end{align}
The two terms are equivalent to the one obtained in the previous section, except for some additional even powers of the two fields.
The field $\phi_i^2$ does not break $\mathcal{Z}_2$ symmetry and therefore scales as $(T-T_c)^0$ at the phase transition. Hence, inserting $\phi_i^2$ into expectation values does not change their dominant scaling behavior at the critical point \cite{zinn-justin_renormalization_2002, cardy1996}\footnote{More precisely, expanding the operator $\phi_i^2$ in a sum of scaling operators, the dominant one (with scaling dimension equal to 0) would be the identity.
}.
The correction at leading order therefore still scales as in \eqref{eq:perturbation_indep_uni}, and the perturbation is relevant whenever the susceptibility diverges.

\section{Constant nonreciprocal coupling on top of reciprocal coupling}
\label{app_constant_nr_unperturbed_r}

In this section, we focus on a quartic potential $V(\phi)= -\frac{a}{2}\phi^2+\frac{b}{4}\phi^4$ for concreteness.
With this choice, the dynamics \eqref{eq:dynevSM}  becomes
\begin{align}
\begin{cases}
    \partial_t\phi_1=\nabla^2\phi_1 + a \phi_1 - b \phi_1^3  + (K_++\delta K_-) \phi_2 + \eta_1\;, \\
    \partial_t\phi_2=\nabla^2\phi_2 + a \phi_2 - b \phi_2^3  + (K_+-\delta K_-) \phi_1 + \eta_2 \;,
\end{cases}
\end{align}

In the presence of a finite reciprocal coupling, we expect the dynamics to simplify if we write in terms of the sum and difference of the two fields. We thus define
\begin{align}
    \psi =& \frac{\phi_1+\phi_2}{\sqrt{2}}\;, &
    \varphi =&\frac{\phi_1-\phi_2}{\sqrt{2}}\;.
\end{align}
Rewriting the dynamical equations in terms of $\psi$ and $\varphi$ yields
\begin{align}
\label{eq:psi_evolution}
    \partial_t\psi=&\nabla^2 \psi +(a+K_+)\psi- b\left(\frac{\psi^3}{2}+\frac{3}{2}\psi\varphi^2\right)-\delta K_-\varphi+\eta_\psi\\
    \label{eq:varphi_evolution}
    \partial_t\varphi=&\nabla^2 \varphi +(a-K_+)\varphi- b\left(\frac{\varphi^3}{2}+\frac{3}{2}\varphi\psi^2\right)+\delta K_-\psi+\eta_\varphi
\end{align}
where $\eta_\psi$ and $\eta_\varphi$ are Gaussian white noises with the same statistics as $\eta_1$ and $\eta_2$.
When $\delta K_-=0$, $\psi$ and $\varphi$ are only coupled through a cubic term.
Neglecting this higher order coupling, we have two uncoupled $\phi^4$ theories that only differ for their linear term, leading to two different transition points.
$\psi$ has a larger linear term, therefore it is the first to undergo the $\mathds{Z}_2$-symmetry-breaking phase transition, as expected.
The cubic term that couples $\psi$ to $\varphi^2$ is irrelevant for the critical properties of the system, since we already know it has to fall in the Ising universality class.
When $\psi$ undergoes the phase transition, $\varphi$ is still subcritical: as such its fluctuations do not exhibit long range correlations and therefore does not change the large scale properties of the system.

The system is therefore equivalent to two field theories with different critical points perturbed by an antisymmetrical coupling. As explained in the main text, the  perturbation $\delta K_-$ is irrelevant in this case.

\section{Random nonreciprocal couplings}
\label{random_nr}
In this section, we compute the variance of the  correction to the order parameter $\langle \delta \phi_i \rangle$ when the perturbation $\delta K_-(\bx)$ is random.
Using Eq. \eqref{eq:perturbation_indep}, we obtain to first order
\begin{align}
\begin{split}
    \overline{\delta \langle\phi_j(\bx,t)\rangle^2} = & \overline{\left( -\epsilon_{ji} \int \delta K_-(\bx')\frac{\delta \langle \phi_j (\bx,t)\rangle_0}{\delta h_j(\bx',t')}\bigg|_{h_j=0}\langle \phi_i(\bx',t')\rangle_0 d\bx' dt' \right)^2}  \\
    =& \int d\bx' d\bx'' dt' dt'' \overline{\delta K_-(\bx')\delta K_-(\bx'')}\frac{\delta \langle \phi_j (\bx,t)\rangle_0}{\delta h_j(\bx',t')}\bigg|_{h_j=0}\langle \phi_i(\bx',t')\rangle_0\frac{\delta \langle \phi_j (\bx,t)\rangle_0}{\delta h_j(\bx'',t'')}\bigg|_{h_j=0}\langle \phi_i(\bx'',t'')\rangle_0\\
    =&\delta \sigma_K^2 \langle \phi_i\rangle_0^2\int d\bx' dt' dt'' \frac{\delta \langle \phi_j (\bx,t)\rangle_0}{\delta h_j(\bx',t')}\bigg|_{h_j=0}\frac{\delta \langle \phi_j (\bx,t)\rangle_0}{\delta h_j(\bx'',t'')}\bigg|_{h_j=0}\;.
\end{split}
\label{eq:computationsrandomcoupling}
\end{align}

Close to the critical point, the response function scales as \cite{cardy1996}
\begin{align}
        \left\langle \frac{\partial \phi_i(x,t)}{\partial h_i(x', t')} \right\rangle_0=\xi^{-d-z+\frac{\gamma}{\nu} } f\left(\frac{x-x'}{\xi}, \frac{t-t'}{\tau}\right)\;.
    \end{align}
Inserting such scaling in the integral, we obtain
\begin{align}
\begin{split}
    \overline{\delta \langle\phi_j(\bx,t)\rangle^2} =& \delta \sigma_K^2 \langle \phi_i\rangle_0^2 \xi^{2\left(-d-z+\frac{\gamma}{\nu} \right)} \int d\bx' dt' dt'' f\left(\frac{x-x'}{\xi}, \frac{t-t'}{\tau}\right)f\left(\frac{x-x''}{\xi}, \frac{t-t''}{\tau}\right)\\
    \propto& \delta \sigma_K^2 \langle \phi_i\rangle_0^2 \xi^{2\left(-d-z+\frac{\gamma}{\nu} \right)}\xi^{d+2 z}\sim \delta \sigma_K^2 |T-T_c|^{2\beta} |T-T_c|^{-2 \gamma +d\nu}
\end{split}
\end{align}

Our computation can be generalized to interactions that are not fully antisymmetric, i.e. to the case in which the system is perturbed by two random interactions coefficients $\delta K_{12}(\bx)$ (for the effect of $\phi_2$ on $\phi_1$) and $ \delta K_{21}(\bx)$ (for the effect of $\phi_1$ on $\phi_2$) such that
\begin{align}
        \overline{\delta K_{12}(x)\delta K_{12}(x')}&=\overline{\delta K_{21}(x)\delta K_{21}(x')}=\delta \sigma_K^2 \delta(x-x')\;, &
        \overline{\delta K_{12}(x)\delta K_{21}(x')}&=\rho \delta \sigma_K^2 \delta(x-x')\;,
    \end{align}
where $\rho$ is a generic correlation coefficient.
We find that the scaling of the correction is unchanged for any value of $\rho$, including in the case of symmetric interactions, for which it matches the equilibrium result.

\section{Nonreciprocally coupled fields with random perturbation}
\label{app:nonrec_with_quenched}
In this section, we consider nonreciprocally coupled fields perturbed by random nonreciprocal interactions
\begin{align}
\label{eq:nonrec_coupling_constant_plus_random}
    K_{ij}=(K_- + \delta K_- (\bx))\epsilon_{ij}\;.
\end{align}
Even though we could directly apply the Harris criterion using our generalization to nonequilibrium systems, hereafter we derive the result in this particular setting.
Lowering the temperature or decreasing $K_-$ the system undergoes a transition from a disordered to a ``swap'' phase with sustained oscillations \cite{Avni2025}.
The order parameter is the angular momentum $\cL$, defined as
    \begin{align}
    \label{eq:definition_momentum}
        \mathcal{L} =\langle \dot\phi_1 \phi_2-\dot \phi_2\phi_1\rangle \equiv \langle \mathcal{O}_L(\bx, t) \rangle\;.
    \end{align}
The scaling of $\cL$ at the transition defines the critical exponent $\beta$ as:
\begin{align}
\label{eq:scaling_momentum_transition}
        \mathcal{L}\sim|K_--K_c(T)|^\beta\;,
    \end{align}
where $K_c(T)$ is the critical line.
The phase transition is believed to fall in the $XY$ universality class \cite{Risler2004,Risler2005,Daviet2024, Zelle2024}, and
the static critical exponents measured in Monte-Carlo simulations agree with their corresponding $XY$ values within uncertainty \cite{Avni2025} (see Figure \ref{fig:CriticalExponents}).

Our aim is to determine if this transition to oscillations is affected by the introduction of the random, inhomogeneous and nonreciprocal perturbation $\delta K_-(\bx)$.
Note that in contrast to the previous sections, now the system is nonreciprocal even before the introduction of the perturbation, so that we will not be able to use equilibrium properties when averaging with respect to $\cS_0$.
The perturbation of the action $\delta \cS$ takes the same form as in the previous section, whereas the unperturbed one has an additional term describing the uniform part of the nonreciprocal interactions.
Defining the operator $\cO_p(\bx',t')=\hat{\phi}_1(\bx)\phi_2(\bx)-\hat{\phi}_2(\bx)\phi_1(\bx)$, we can express the action as
\begin{align}
\cS_0(\hat{\phi}_1,\hat{\phi}_2,\phi_1,\phi_2)=&\cS_0(\hat{\phi}_1,\phi_1,h_1) + \cS_0(\hat{\phi}_2,\phi_2,h_2) + K_-\cO_p\;,  &
\delta \cS(\hat{\phi}_1,\hat{\phi}_2,\phi_1,\phi_2)=&\delta K_-(\bx)\cO_p\;.
\end{align}
To linear order in $\delta K_-$, the deviation of the order parameter is given by
\begin{align}
    \delta \cL = \delta \langle \cO_{L}(\bx,t)\rangle= -\langle\int_{\bx', t'} \cO_{L}(\bx,t)\delta K_-(\bx')\cO_p(\bx', t') \rangle_0\;,
\end{align}
where $\langle \cdot \rangle_0$ indicates averaging with respect to $\cS_0$.
The first order correction averages to 0: we therefore compute its variance $\overline{\delta \cL^2}$ as
\begin{align}
\label{eq:variance_deviation_nonrec_nonrec_random}
    \overline{\delta \langle\cO_L(\bx,t)\rangle^2} =& \int d\bx dt^{\prime} dt^{\prime\prime} \delta \sigma_{K}^2 \langle \cO_{L}(\bx,t)\cO_p(\bx',t')\rangle_0 \langle \cO_{L}(\bx,t) \cO_p(\bx',t'^{\prime})  \rangle_0\;.
\end{align}
To obtain the scaling of the correlators in the integrand at the transition, we remark that
\begin{align}
\label{eq:scaling_for_op}
    \int d\bx dt^{\prime} \langle \cO_{L}(\bx,t) \cO_p(\bx',t')  \rangle_0 = \frac{\delta\langle \cO_{L}(\bx,t)\rangle_0}{\delta K_-}\bigg|_{K_-=K_-}\;.
\end{align}
As $\langle \cO_{L}(\bx,t) \rangle_0 \sim (K_- -K_c)^{\beta}$ close to the transition, we deduce that
\begin{align}
    \frac{\delta\langle \cO_{L}(\bx,t)\rangle_0}{\delta K_-}\bigg|_{K_-=K_-}\sim (K_--K_c)^{\beta-1}\;.
\end{align}
We further assume that $\langle \cO_{L}(\bx,t) \cO_p(\bx',t')  \rangle_0 \sim \xi^{-\alpha}$, with $\xi$ being the correlation length scaling as $\xi\sim(K_--K_c)^{-\nu}$. The left hand side of \eqref{eq:scaling_for_op} then scales as
\begin{align}
\label{eq:scaling_int_op}
    \int d\bx dt^{\prime} \langle \cO_{L}(\bx,t) \cO_p(\bx',t')  \rangle_0 \sim \xi^{d+z-\alpha} \sim (K_--K_c)^{\nu(\alpha-d-z)}\;.
\end{align}
Equating scaling \eqref{eq:scaling_int_op} with the scaling of the right hand side of \eqref{eq:scaling_for_op}, we finally obtain $\alpha$ as
\begin{align}
    \nu(\alpha -d -z)= \beta -1 \Longrightarrow \alpha = d+z+\frac{\beta -1}{\nu}\;.
\end{align}
Now that the critical behavior of $\langle \cO_{L}(\bx,t) \cO_p(\bx',t')  \rangle_0$ is determined, we deduce the scaling of $\overline{\delta \langle\cO_L(\bx,t)\rangle^2}$ using \eqref{eq:variance_deviation_nonrec_nonrec_random} as
\begin{align}
    \overline{\delta \langle\cO_{L}(\bx,t)\rangle^2} \sim \xi^{-2\alpha +2z+d} \sim (K_--K_c)^{2(\beta-1)+\nu d}\;,
\end{align}
Comparing the standard deviation of the correction to the unperturbed order parameter, we find the disorder to be relevant if $\nu d<2$,
which corresponds to the standard Harris criterion.
Note that our computation can be straightforwardly generalized to an $O(N)$ model, in which each field is replaced by a vector with $N$ components.
Promoting the fields to $N$-dimensional vectors and the products in $\mathcal{O}_L$ and $\mathcal{O}_P$ to scalar products, the same derivation carries through and leads to the same result.
\section{Critical exponents calculation from Monte-Carlo simulations}
\label{mcmc}
Figure 2 of the main text shows the critical exponents $\nu$, $\gamma$, and $\beta$, as well as the Binder cumulant at the critical point $U^*\equiv U(T=T_c)$~\cite{Binder1981} of the 3D NR Ising model in two cases with distinct couplings. The case where $K_+=0$ and $\delta K_-=0.1$ (with $k_BT=1$) is taken from Ref.~\cite{Avni2025a}, while the case with $K_+=0.5$ and $\delta K_-=0.1$ (with $k_BT=1$) results from new simulations.
These exponents and $U^*$ are obtained from finite-size scaling analysis using the same procedure as detailed in \textsection VIII A of Ref.~\cite{Avni2025a}. In Fig. 4, we show the finite-size scaling numerics, which should be compared with Fig.~14 of Ref.~\cite{Avni2025a}. Note that, as in Ref.~\cite{Avni2025a}, we use the order parameter
\begin{equation}
    R = \langle s\rangle \equiv \left\langle  \sqrt{(M_1^2+M_2^2)/2} \right\rangle,
\end{equation}
where $M_1$ and $M_2$ are the total magnetizations of species 1 and 2 respectively, although other choices, such as $\langle M_1+M_2\rangle$, are also possible. The susceptibility $\chi$, and the Binder cumulant $U$ shown in Fig. 2 are defined as
\begin{align}
\chi=L^{d}\left(\langle s^{2}\rangle-\langle s\rangle^{2}\right)
\qquad
\text{and}
\qquad
U = 1-\frac{\langle s^4\rangle }{3\langle s^2\rangle^2}\;,
\end{align}
where $L$ is the linear system size and $d$ the dimension. The parameter that represents the distance from the critical point in our calculation is $\tilde{J} = 2 d J/(k_B T)$, with $J$ the coupling between nearest-neighbors spins.

The explicit values we obtain for the critical exponents (summarized in Fig. 2) are
\begin{align}
    \nu&=0.675\pm0.005 &
    \gamma&=1.328\pm0.009 &
    \beta&=0.347\pm0.002\
\end{align}
for $K_+ =0$ and $\delta K_- = 0.1$ ~\cite{Avni2025a}
and
\begin{align}
    \nu&=0.631\pm0.004 &
    \gamma&=1.253\pm0.009 &
    \beta&=0.318\pm0.002
\end{align}
for $K_+ =0.5$ and $\delta K_- = 0.1$.
The Ising and XY values are respectively $\nu_{I}=0.630$, $\gamma_{I}=1.237$, $\beta_{I}=0.326$~\cite{Pelissetto2002} and $\nu_{XY}=0.672$, $\gamma_{XY}=1.318$, $\beta_{XY}=0.349$~\cite{Pelissetto2002,Campostrini2001}, respectively.

We note a significant discrepancy between Ising's and nonreciprocal Ising's $\beta$ when $K_+ =0.5$ and $\delta K_- = 0.1$ (approximately four standard deviation, see also Fig. 2).
Note, however, that the reported standard deviation reflects only statistical uncertainties due to finite sampling (estimated via bootstrapping), and does not take into account other systematic errors such as finite-size corrections~\cite{binder2022monte}.
Repeating the same exact procedure for deriving the critical exponents while excluding the smallest system size ($L=20$), we obtain the new values for the critical exponents as
\begin{align}
    \nu&=0.636 &
    \gamma&=1.262 &
    \beta&=0.323
\end{align}
in which $\beta$ is much closer to the Ising exponent $\beta_I$.
We conclude that the value of $\beta$ is compatible with the Ising universality class if we account for finite-size uncertainties on top of finite sampling errors.
\begin{figure*}
    \centering
    \includegraphics[width=1\linewidth]{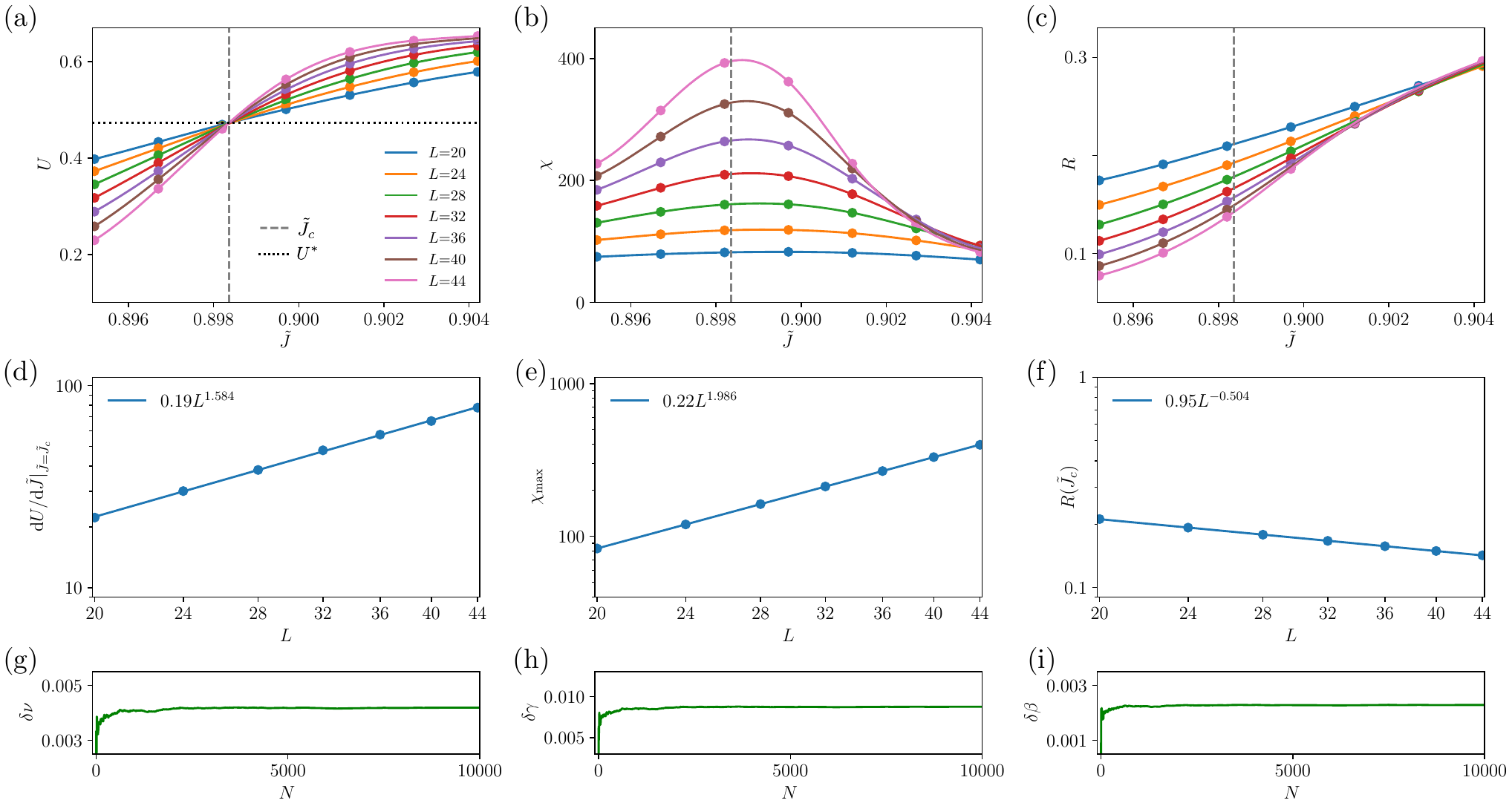}
    \caption{Numerical determination of the critical exponents in the 3D NR Ising model with $K_+=0.5$ and $\delta K_-=0.1$. The procedure, as well as the error evaluation, is similar to Fig.~14 in Ref.~\cite{Avni2025a}.}
\label{fig:CriticalExponents_appendix}
\end{figure*}

\section{An example of nonrelaxational unperturbed dynamics}
\label{example_nonrelaxational}
In this section we discuss the generalization of our main result on the relevance of a uniform nonreciprocal coupling to the case in which the two fields have a nontrivial Poisson-bracket relation with an auxiliary conserved density.
To do so, we consider as the unperturbed system two uncoupled copies of model F, a nonrelaxational model which was introduced to describe the critical dynamics of superfluid helium \cite{hohenberg1977,Folk2006}, and we add a nonreciprocal coupling.
Note that our choice of couplings is arbitrary; it consists in having each order parameter act as an external field for the other, with opposite prefactors.
Model F involves a complex field which is invariant under phase shifts ($U(1)$ symmetry), and a conserved density which can be interpreted as the conjugate variable (in terms of Poisson brackets) to the phase of the field.
The density is only conserved in the absence of an external field \cite{halperin_renormalization-group_1976}, and our nonreciprocal perturbation also break this conservation law.

The free-energy functional of Model F can be written as:
\begin{align}
    F_0[\phi_i, m_i] = \int _{\bx} V(\phi_i) + \frac{1}{2}|\nabla \phi_i|^2 +\gamma_0 |\phi_i|^2 m_i+ \frac{1}{2C}m_i^2 - h_i^m m_i- \text{Re}(h_i\phi_i^*)
\end{align}
A natural way to introduce nonreciprocity in the model is by defining two ``selfish'' free-energy functionals:
\begin{align}
    F_1[\phi_1, \phi_2, m_1, m_2]= F_0[\phi_1, m_1] - \delta K \int _{\bx} \text{Re}(\phi_1^* \phi_2)\\
    F_2[\phi_1, \phi_2, m_1, m_2]= F_0[\phi_2, m_2] + \delta K \int _{\bx} \text{Re}(\phi_1^* \phi_2)
\end{align}
These definitions generalize the idea that $\phi_1$ tends to align with $\phi_2$, while $\phi_2$ tends to antialign with $\phi_1$. $\delta K$ is again the (infinitesimal) strength of the nonreciprocal perturbation.
From the free-energy functionnals, we deduce the dynamics as
\begin{align}
    \partial_t \phi_i &= - 2 \Gamma \frac{\delta F_i}{\delta \phi_i^*} - i g \frac{\delta F_i}{\delta m_i} +\eta_i \\
    \partial_t m_i &= \lambda \nabla^2 \frac{\delta F_i}{\delta m_i} + 2 g \text{Im}\left( \phi^*_i\frac{\delta F_i}{\delta \phi_i^*}\right) + \nabla \cdot \zeta_i\\
    \langle \eta_i(\bx, t)\eta_j^*(\bx', t')\rangle&= 4 T \text{Re}(\Gamma)\delta_{ij}\delta(\bx-\bx')\delta(t-t')\\
    \langle \zeta_i(\bx, t)\cdot \zeta_j(\bx', t')\rangle&= 2T \lambda\delta_{ij} \delta(\bx-\bx')\delta(t-t')\;,
\end{align}
where $\Gamma$ can be complex and $g$ represents a nondissipative coupling.
This system can also be treated via the MSRDJ formalism, yielding the unperturbed action as
\begin{align}
    \cS_0= \sum_i \left(\partial_t \phi_i + 2 \Gamma \frac{\delta F_0}{\delta \phi_i^*} + i g \frac{\delta F_0}{\delta m_i}\right)\hat \phi_i + \partial_t m_i\hat m_i -\left(\lambda \nabla^2 \frac{\delta F_0}{\delta m_i} + 2 g \text{Im}\left( \phi^*_i\frac{\delta F_0}{\delta \phi_i^*}\right) \right)\nabla^2\hat m_i + \frac{T}{2}\left(4\text{Re}(\Gamma)\hat \phi_i^2 +2\lambda (\nabla\hat m_i)^2\right)\nonumber\;,
\end{align}
and the perturbation as
\begin{align}
    \delta \cS = \delta K \left( \Gamma \left(\phi_2 \hat\phi_1 - \phi_1 \hat\phi_2 \right) - g \left(\text{Im}(\phi_1^*\phi_2)\nabla^2\hat m_1-\text{Im}(\phi_2^*\phi_1)\nabla^2\hat m_2  \right)\right)\;.
\end{align}
Instead of working with the real and imaginary part, it is convenient to treat $\phi_i$ and $\phi_i^*$ as independent; the same can be done for the perturbation fields $h_i$.
The response function of the unperturbed system can then be expressed as a $2\times 2$ matrix reading
\begin{align}
\hat R(\bx, \bx', t, t')=\begin{pmatrix}
    \displaystyle\frac{\delta \langle\phi_i(\bx, t)\rangle_0}{\delta h_i(\bx', t)}  & \displaystyle\frac{\delta \langle\phi_i(\bx, t)\rangle_0}{\delta h_i^*(\bx', t)}\\
    \displaystyle\frac{\delta \langle\phi_i^*(\bx, t)\rangle_0}{\delta h_i(\bx', t)}  & \displaystyle\frac{\delta \langle\phi_i^*(\bx, t)\rangle_0}{\delta h_i^*(\bx', t)}
\end{pmatrix}
\end{align}
The matrix elements of $\hat R(\bx, \bx', t, t')$ can be obtained by inserting $F_0$ inside the action, and computing and taking derivatives with respect to $h$ and $h^*$ as
\begin{align}
\label{eq:responseFmatrix}
    \frac{\delta \langle\phi_1(\bx, t)\rangle_0}{\delta h_1(\bx', t)} &= \left\langle \phi_1(\bx,t)\left( \Gamma \hat\phi_1  - \frac{g}{2} \phi_1^*\nabla^2\hat m_1\right)|_{\bx', t'}   \right\rangle_0,\\
    \frac{\delta \langle\phi_1(\bx, t)\rangle_0}{\delta h_1^*(\bx', t)} &= \left\langle \phi_1(\bx,t) \frac{g}{2} \phi_1\nabla^2\hat m_1|_{\bx', t'} \right\rangle_0.
\end{align}
The other two response functions can similarly be obtained by complex conjugation.
The matrix structure of the response function reflects that the order parameter, being complex, has a $O(2)$ symmetry, and responds differently to perturbations parallel and perpendicular to itself.
The perturbation to the order parameter can be written as:
\begin{align}
    \delta \langle \phi_1(\bx, t) \rangle &= -  \left\langle \phi_1(\bx,t) \int_{\bx' t'} \delta \cS|_{\bx', t'}  \right\rangle_0 \nonumber \\&= - \delta K \left\langle \phi_1(\bx,t) \int_{\bx' t'} \left( \Gamma \left(\phi_2 \hat\phi_1 - \phi_1 \hat\phi_2 \right) - g \left(\text{Im}(\phi_1^*\phi_2)\nabla^2\hat m_1-\text{Im}(\phi_2^*\phi_1)\nabla^2\hat m_2  \right)\right)|_{\bx', t'}   \right\rangle_0 \\
    &=- \delta K\left( \langle \phi_2\rangle_0 \int_{\bx' t'} \left\langle \phi_1(\bx,t)\left( \Gamma \hat\phi_1  - \frac{g}{2} \phi_1^*\nabla^2\hat m_1\right)|_{\bx', t'} \right\rangle_0+\langle \phi_2^*\rangle_0 \int_{\bx' t'} \left\langle \phi_1(\bx,t) \frac{g}{2} \phi_1\nabla^2\hat m_1|_{\bx', t'} \right\rangle_0 \right)\nonumber
\end{align}
We can jointly express the correction to $\phi_1$ and its complex conjugate as:
\begin{align}
    \delta \begin{pmatrix}
        \langle \phi_1(\bx, t)\rangle_0\\
        \langle \phi_1^*(\bx, t)\rangle_0
        \end{pmatrix} = - \delta K  \int_{\bx' t'}\hat  R(\bx, \bx', t, t') \begin{pmatrix}
        \langle \phi_2\rangle_0\\
        \langle \phi_2^*\rangle_0
    \end{pmatrix}= - \delta K \hat  \chi\begin{pmatrix}
        \langle \phi_2\rangle_0\\
        \langle \phi_2^*\rangle_0
        \end{pmatrix}
\end{align}
where
\begin{align}
    \hat  \chi= \begin{pmatrix}
    \displaystyle\frac{\delta \langle\phi_i\rangle_0}{\delta h_i}  & \displaystyle\frac{\delta \langle\phi_i\rangle_0}{\delta h_i^*}\\
    \displaystyle\frac{\delta \langle\phi_i^*\rangle_0}{\delta h_i}  & \displaystyle\frac{\delta \langle\phi_i^*\rangle_0}{\delta h_i^*}
\end{pmatrix}
\end{align}
is the static susceptibility, obtained by integrating $\hat R$ over time and space.
The eigenvalues of $\hat \chi$ are the susceptibility to parallel and perpendicular perturbations, for which we can define two different critical exponents $\chi_\parallel\sim (T-T_c)^{-\gamma_\parallel}$ and $\chi_\perp\sim (T-T_c)^{-\gamma_\perp}$.
Nonreciprocity is a relevant perturbation whenever either of the static susceptibilities diverges, i.e. for $\max(\gamma_\parallel, \gamma_\perp)>0$.
Both $\gamma_\parallel$ and $\gamma_\perp$ are positive for equilibrium systems.
Other choices are possible to introduce nonreciprocity, which may lead to different results (a calculation is required for each of them).
For example, we could have included a nonreciprocal force only in the equations for the $\phi_i$. This would lead to the combination of different response functions, whose evaluation could be nontrivial.

\section{Role of field inversion symmetries in our procedure}
\label{app:DP}
In this section, we illustrate the importance of inversion symmetries on which our procedure relies. To this aim, we study a system lacking such symmetries: Directed Percolation (DP), a widely studied model describing spreading phenomena ranging from forest fires to epidemics, which undergoes a nonequilibrium phase transition between an active and an inactive phase \cite{hinrichsen2000,Odor2004,Henkel2008,Henkel2010,Sieberer2025}.
The corresponding microscopic processes can be coarse-grained into an effective  Langevin equation for the macroscopic density of particles, given by \cite{hinrichsen2000,janssen_nonequilibrium_1981, cardy_directed_1980}:
\begin{align}
\label{eq:directed_perco_single}
    \partial_t\rho&=\nabla^2\rho +m\rho -\lambda \rho^2 + \sqrt{\rho}\eta\;,
\end{align}
where $\eta$ is a white Gaussian noise of amplitude $T$ while $m$ and $\lambda$ are fixed constants related to creation and pairwise destruction rates.

A natural way to introduce nonreciprocal interactions between two species described by \eqref{eq:directed_perco_single} is via a quadratic term: the two species interact only if both are present.
The model becomes then equivalent to a Lotka-Volterra system with space and demographic fluctuations \cite{janssen_spontaneous_1997,tauber_stochastic_2024, de_giuli_dynamical_2022, arnoulx_de_pirey_self-organized_2025}, which is given by
\begin{align}
    \partial_t\rho_1&=\nabla^2\rho_1 +m\rho_1 -\lambda \rho_1^2 + \sqrt{\rho_1}\eta_1 + \delta K_-\rho_1\rho_2 \;, \\
    \partial_t\rho_2&=\nabla^2\rho_2 +m\rho_2 -\lambda \rho_2^2 + \sqrt{\rho_2}\eta_2 -\delta K_-\rho_1\rho_2 \;.
\end{align}
Note a crucial difference with respect to the case studied in the main text: the $\rho_i$ are densities, and as such must be positive.
This means that the system is not symmetric under the inversion of the two fields: an exchange between the identities of the two species followed by a reversion of one of the two fields does not necessarily leave the action invariant.
Therefore, the symmetry enforcing an even behavior of the critical temperature as a function of $\delta K_-$ is broken.
We can no longer rule out that species 1 develops a critical temperature that differs from the one of species 2.
Indeed, we expect the species unfavoured by the interaction (the `prey') to encounter the transition to the absorbing phase before than it would have in the unperturbed case.
Following this transition, the `predator' undergoes a transition similar to the unperturbed case since the interaction term vanishes in the absence the other species.
The irrelevance of the perturbation is confirmed by the RG results of reference \cite{janssen_spontaneous_1997}.

\end{document}